\documentclass[sigconf]{acmart}

\AtBeginDocument{%
  \providecommand\BibTeX{{%
    \normalfont B\kern-0.5em{\scshape i\kern-0.25em b}\kern-0.8em\TeX}}}

\setcopyright{acmcopyright}
\copyrightyear{2023}
\acmYear{2023}
\acmDOI{XXXXXXX.XXXXXXX}

\acmConference[WSDM Cup'23]{the Sixteenth ACM International Conference on Web Search and Data Mining}{February 27-March 3,
  2023}{Singapore, Singapore}
\acmPrice{15.00}
\acmISBN{978-1-4503-XXXX-X/18/06}

\usepackage{enumitem}
\usepackage{balance}
\usepackage{booktabs}
\usepackage{threeparttable}
\begin{document}

\title{Feature-Enhanced Network with Hybrid Debiasing Strategies for Unbiased Learning to Rank}

\author{Lulu Yu}
\authornotemark[1]
\thanks{*Equal contribution, and the order is determined randomly.}

\affiliation{
 \institution{CAS Key Lab of Network Data Science and Technology, ICT, CAS}
 \institution{University of Chinese Academy of Sciences}
 \city{Beijing}
 \country{China}
}
\email{nothing_0_1@163.com}

\author{Yiting Wang}
\authornotemark[1]
\affiliation{
 \institution{CAS Key Lab of Network Data Science and Technology, ICT, CAS}
 \institution{University of Chinese Academy of Sciences}
 \city{Beijing}
 \country{China}
}
\email{wangyiting21s@ict.ac.cn}

\author{Xiaojie Sun}
\authornotemark[1]
\affiliation{
 \institution{CAS Key Lab of Network Data Science and Technology, ICT, CAS}
 \institution{University of Chinese Academy of Sciences}
 \city{Beijing}
 \country{China}
}
\email{sunxiaojie21s@ict.ac.cn}

\author{Keping Bi}
\affiliation{
 \institution{CAS Key Lab of Network Data Science and Technology, ICT, CAS}
 \institution{University of Chinese Academy of Sciences}
 \city{Beijing}
 \country{China}
}
\email{bikeping@ict.ac.cn}

\author{Jiafeng Guo}
\affiliation{
 \institution{CAS Key Lab of Network Data Science and Technology, ICT, CAS}
 \institution{University of Chinese Academy of Sciences}
 \city{Beijing}
 \country{China}
}
\email{guojiafeng@ict.ac.cn}

\begin{abstract}
Unbiased learning to rank (ULTR) aims to mitigate various biases existing in user clicks, such as position bias, trust bias, presentation bias, and learn an effective ranker. In this paper, we introduce our winning approach for the ``Unbiased Learning to Rank'' task in WSDM Cup 2023. We find that the provided data is severely biased so neural models trained directly with the top 10 results with click information are unsatisfactory. So we extract multiple heuristic-based features for multi-fields of the results, adjust the click labels, add true negatives, and re-weight the samples during model training. Since the propensities learned by existing ULTR methods are not decreasing w.r.t. positions, we also calibrate the propensities according to the click ratios and ensemble the models trained in two different ways. Our method won the 3rd prize with a DCG@10 score of 9.80, which is 1.1\% worse than the 2nd and 25.3\% higher than the 4th. 
\end{abstract}



\maketitle

\section{Introduction}

Learning to Rank (LTR) has been playing an essential role in a wide variety of real-world systems, especially search engines. It typically aims to learn a scoring function of various types of features extracted from a query, document, and their matching, such as location, quality, and relevance, towards the document label corresponding to the query. Since user clicks on a document indicate users' implicit feedback on the document's relevance, and it is cheap to collect a large scale of them, a common practice in search engines is to train an LTR model with click data. Whether a document is clicked and how long it has been browsed (i.e., dwelling time) can act as relevance signals.

Despite their effectiveness, such signals contain a lot of noise and biases. They may be not accurate to indicate relevance and are susceptible to factors such as display position, document length. Thus, optimizing the model directly toward the click signals could lead to unsatisfied performance. Aware of this issue, unbiased learning to rank (ULTR) \cite{ai2021unbiased, ai2018unbiased} has attracted much attention from the research community. Most ULTR methods are proposed to address position bias \cite{ai2018unbiased}, selection bias \cite{cai2022hard}, and trust bias \cite{trustPBM2019addressing}. 

ULTR models have been shown to be effective on data synthesized according to position-based click model assumptions \cite{ai2018unbiased,ai2021unbiased} from public datasets such as Yahoo! \cite{chapelle2011yahoo} and LETOR \cite{qin2010letor}. However, real-world user click behaviors are much more complex and ULTR methods may not consistently achieve good performance on real data. Based on the search logs of the largest Chinese search engine, Baidu, the WSDM Cup 2023 presents a task of ``Unbiased Learning to Rank'' to alleviate the bias in real-world click data. Participants need to learn an unbiased ranker from the click data and evaluate their model on a hidden human-annotated test set. 

In this competition, we adopt multiple strategies to mitigate multiple biases in the click data during training. Our solution mainly includes several strategies: 1) We extract multiple word-based matching features across multi-fields of documents and incorporate them with the transformer-based model for ULTR. 2) We conduct careful negative sampling and sample re-weighting to alleviate the false negative issue in the non-clicked documents and also bridge the gap between training and test data. 3) We estimate the propensity values according to the click ratios at each position and adjust the learning target by considering both clicks and a well-performing unbiased matching feature. 

We find that traditional word-based matching features outperform the neural model trained with the click data by a large margin. Adding random negative samples will boost the model performance. The propensity values learned by ULTR methods such as DLA \cite{ai2018unbiased} are not decreasing with respect to positions, which is weird and inconsistent with existing studies. Adjusting the propensities according to click ratios benefits the model training. Our team ranked 3rd and our performance is only 1.1\% lower than the 2nd and 25.3\% higher than the 4th.

\begin{table}[htbp]
  \caption{A summary of the notations used in this paper. }
  \label{tab:notations}
  \begin{tabular}{cl}
    \toprule
    Notation &Meaning\\
    \midrule
    $w$ & word of one query\\
    $t$ & one text\\
    $q$ & one query\\
    $\pi_q$ & the candidate document list of query $q$ \\
    $C$ & one texts collection\\
    $N$ & the number of texts in $C$\\
    $p_s(w|t)$ & the probability of seen word $w$ occurs in text $t$\\
    $L_t$ & the length of one text $t$\\
    $L_{avg}$ & the average length of texts in collection\\ 
    $df(x)$ & document frequency of term $x$\\
    $c(w;t)$ & count of $w$ exists in one text $t$\\
    $|t|_u$ & the number of unique terms in text $t$\\
    $|t|$ & the total count of words in text $t$\\
    $c_{ij}$ & click label of $i^{th}$ query's $j^{th}$ sample\\
    $f_{ij}$ & feature label of $i^{th}$ query's $j^{th}$ sample\\
    $l_{ij}$ & final label of $i^{th}$ query's $j^{th}$ sample\\
    $cr_i$ & click ratio of $i^{th}$ ranking position\\
    $pw_i$ & propensity weight of $i^{th}$ ranking position\\
  \bottomrule
\end{tabular}
\end{table}

\section{Training Data Preparation}
In this section, we describe our procedure of data preparation for training. 
We pre-process the data first and then extract the representative features, select negative samples, and adjust the label.

\subsection{Data Pre-processing}
Due to the biased and noisy characteristics of the search logs, it could be beneficial to pre-process and clean the raw data before the training procedure. In this task, there are around 1.2 billion query-document pairs in 2,000 partitions, which is extremely large and it will take a long training time and large computation costs to train on the whole dataset. Due to the limitation of our computing resources, we only use one partition to train the model and leave the research of training on larger data with our strategy for the future. Besides, we found that there exist some search sessions with no clicks on the whole candidate list and these samples do not contribute to the training process. 
So we discard such queries. What's more, we filter out the queries whose candidate documents are fewer than 10 since such queries are too rare and unpopular to have related resources or the logging may be incomplete. 

\subsection{Feature Extraction}
In order to enhance the representation ability of the ranking model, we extract traditional word-based exact matching features and incorporate them in the Transformer. We extract well-known features such as BM25 \cite{robertson1994some} and query likelihood with different smoothing \cite{zhai2004study} methods for multiple document fields, i.e., title, snippet, and the entire content. These features are projected to hidden space and then combined with the vector output of [CLS] by the Transformer layers and go through MLP layers to produce the final matching score. We conduct Gaussian normalization to convert the features to a similar scale. The features are calculated as follows:

\begin{itemize}[leftmargin=*]

\item \textbf{TF:} The average frequency of query terms in the title, snippet, and both.
\item \textbf{IDF:} The sum inverse document frequency of query terms in the title, snippet, and both. The IDF of one word in one collection is computed as the following: 

\begin{equation}
    IDF(w,C) = log(\frac{N-df(w)+0.5}{df(w)+0.5})
\end{equation}

\item \textbf{TF-IDF:} The sum value of TF $\cdot$ IDF of query terms in the title, snippet, and both.
\item \textbf{Length:} The length of the title, snippet, and the entire document.
\item \textbf{BM25:} The scores of BM25 \cite{robertson1994some} on the title, snippet, and both, are calculated by the following formulation.

\begin{equation}
    BM25 = \sum\nolimits_widf(w,C)\cdot\frac{(k_1+1)c(w;t)}{K+c(w;t)}\cdot\frac{(k_2+1)c(w;q)}{k_2+c(w;q)}
\end{equation}

where $K$ denotes $k_1(1-b+b\cdot\frac{L_t}{L_{avg}})$.

\item \textbf{LMJM:} The scores of Language Model (LM) with Jelinek-Mercer (JM) \cite{zhai2004study} on title, snippet, and both. The $p_s(w|t)$ of JM is computed as the following, where $\lambda$ is used to control the influence of each frequency:

\begin{equation}
  {p_s(w|t)}_{JM} = (1-\lambda)p(w|t)+\lambda p(w|C)
\end{equation}

\item \textbf{LMDIR:} The scores of LM with Dirichlet (DIR) \cite{zhai2004study} smoothing on title, snippet and the both. The $p_s(w|t)$ of DIR is computed as the following where $\mu$ is a Bayesian smoothing parameter: 
\begin{equation}
  {p_s(w|t)}_{DIR} = \frac{c(w;t) + \mu p(w|C)}{ \sum\nolimits_w c(w;t)+\mu}
\end{equation}

\item \textbf{LMABS:} The scores of LM with absolute discounting (ABS) \cite{zhai2004study} on title, snippet and the both. The $p_s(w|t)$ of ABS is computed as the following where $\sigma$ is a discount constant:
\begin{equation}
  {p_s(w|t)}_{ABS} = \frac{max(c(w;t)-\sigma, 0)}{\sum\nolimits_w c(w;t)} + \frac{\sigma |t|_u}{|t|}p(w|C)
\end{equation}

\end{itemize}

\begin{figure*}
\centering
\includegraphics[width=0.9\textwidth]{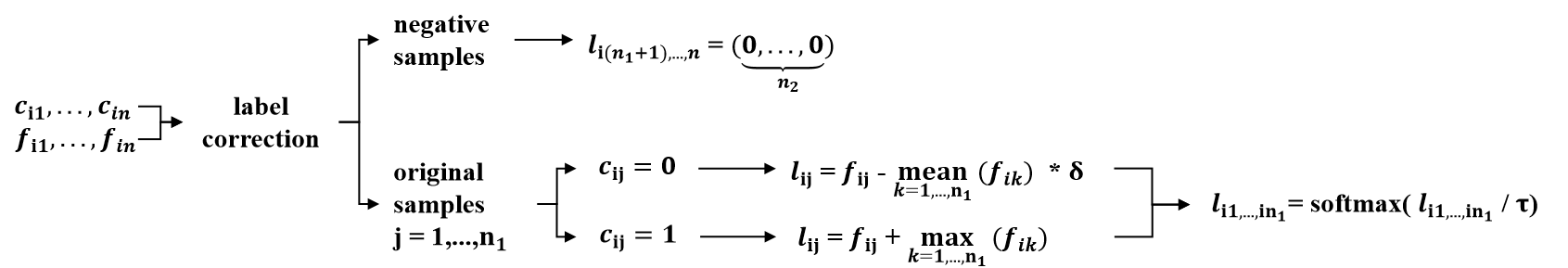}
\caption{The workflow of label adjustment.}
\label{pic:label_correction}
\end{figure*}
\subsection{Negative Sample Selection}


The dataset provided in the unbiased learning to rank task in WSDM Cup 2023 contains a large number of queries, a list of top-10 documents displayed to users under each query, and user feedback on these documents. Since a large majority of search traffic is from high-frequency queries and they usually have sufficient relevant results, the non-clicked documents are likely to be false negatives. Hence, it could harm the capabilities of the model of differentiating relevant documents from irrelevant ones. Aware of this problem, we introduce true negative samples into model training, and the negative samples are randomly selected from the collection of various distinct queries and documents in the same batch.
In addition, the user's last click behavior usually indicates that the user has stopped checking the result list. We think that the documents after the last click position are not examined rather than completely irrelevant, especially for head queries. So we replace the documents that appear after the last clicked document with random negative samples to alleviate the false negative problem.

\subsection{Label Adjustment}
As we found training with click labels does not perform as well as exact matching features alone which indicates that the bias in the click data is huge. Since heuristic-based features are not biased, we consider using the best among them to adjust the click label. For each query's candidate documents, including the original top 10 documents and random negative samples, we utilize their feature values to generate a more learning-friendly label as shown in Equation \eqref{eq:label}. Then apply Softmax with temperature (denoted as $\tau$) to adjust the distribution of the original samples' new labels.
\begin{equation}
\label{eq:label}
l_{ij}=\left\{
\begin{array}{cc}
f_{ij}+reward\_item,&c_{ij}=1\\
f_{ij}-penalty\_item,&c_{ij}=0
\end{array}
\right.
\end{equation}

Simply, we use the mean feature value of the top 10 documents multiplied by one proportion (denoted as $\delta$) as the penalty item. As for reward items, we designed two methods depending on whether to keep the labels of clicked documents larger than the labels of non-clicked ones. One is using the maximum feature value of the top 10 documents, the other is using the same value as the penalty item. Experiments show that the former is better. The label adjustment process is shown in Figure \ref{pic:label_correction}. Here, $n$, $n_1$, and $n_2$ represent the number of total, original, and sampled negative candidate results of each query.

\section{Unbiased Learning}
We utilized several strategies including negative sampling, label adjustment, and sample re-weighting to eliminate the problems caused by bias and noise. 

\subsection{Sample Re-weighting}
In contrast to the training data that mostly consists of high-frequency queries, the test set only contains unique queries and the number of long-tail queries is larger than head queries. Due to this distribution discrepancy, a model trained on the click data may not have a good enough performance on the test set. To address the training distribution bias towards head queries, we introduce a weighting mechanism that decreases the weights of head queries: 
\begin{equation}
  w_q = \frac{\alpha+1} {\beta+freq_q}
\end{equation}
The $\alpha$ and $\beta$ are the hyper-parameters to control the transformation of the original weight (i.e. the original weight equals 1) for each sample, and $freq_q$ denotes the counts of the query $q$. The weight $w_q$ is multiplied by the loss of each sample. 
Using this strategy, the effect of the samples of high-frequency (appearing hundreds and thousands of times or larger) is weakened and the low-frequency samples are enhanced.
We treat the queries with the same anonymized string as the repeated queries in this paper.


\subsection{Inverse Propensity Weighting}
According to the results in \cite{zou2022large}, the dual learning algorithm (DLA) \cite{ai2018unbiased} performs best among the ULTR algorithms. Thus, we first choose DLA as our training strategy and the loss function we used is as follows.
\begin{equation}
L_{ranking}= - \sum\limits_{x\in \pi_q} \frac{P(o_q^1=1 | \pi_q)}{P(o_q^x=1 | \pi_q)} \cdot \log \frac{e^{f(x)}}{\sum\nolimits_{z\in \pi_q} e^{f(z)}}
\end{equation}

\begin{equation}
L_{observation}= - \sum\limits_{x\in \pi_q} \frac{P(r_q^1=1 | \pi_q)}{P(r_q^x=1 | \pi_q)} \cdot \log \frac{e^{g(x)}}{\sum\nolimits_{z\in \pi_q} e^{g(z)}}
\end{equation}

where $f(x)$ and $g(x)$ denote the output of the ranking model and propensity model, respectively. The probabilities of observation and relevance are computed as follows. 
\begin{equation}
    P(r_q^x=1|\pi_q)=\frac{e^{f(x)}}{\sum\nolimits_{z\in \pi_q}e^{f(z)}}
\end{equation}
\begin{equation}
    P(o_q^x=1|\pi_q)=\frac{e^{g(x)}}{\sum\nolimits_{z\in \pi_q}e^{g(z)}}
\end{equation}

DLA jointly learns a propensity model and ranking model to mitigate position bias in the data of user feedback without result randomization. Surprisingly, we found that the propensities learned with DLA and other ULTR methods are not decreasing even with user clicks as labels. The reason may be that the click data of real-world search engines contains not only position bias but also presentation bias, invalid clicks, vicious assault, etc. The learning process of propensity value may be impaired owing to the above-mixed factors. What's more, we construct new labels by combining the clicks and exact matching signals, which makes the situation  even more different from the original DLA. 

In order to ensure the stability of the learning process for the biased data, we calculated the click ratio on different ranking positions to estimate the reasonable inverse propensity weights as the following:
\begin{equation}
    pw_i=(\frac{cr_1}{cr_i})^{\gamma}
\end{equation}
Here, $\gamma$ is used to adjust the relative size of propensity weights. After our experiment, we fix it as 0.25. Then the propensity weight in the fixed training strategy is set as 1, 1.19, 1.44, 1.58, 1.89, 1.85, 1.95, 2.12, 2.26, and 2.51 for the top-10 ranked documents.

Meanwhile, the labels of the non-clicked samples have been adjusted by their feature values and are no longer zero, which means that the attention-based cross-entropy loss also includes such samples. The inverse propensity weights of such non-clicked documents are set to 1. Using this technique, the loss function becomes:
\begin{equation}
    L_{fixed}= - \sum\limits_{x\in \pi_q} pw_x \cdot \log \frac{e^{f(x)}}{\sum\nolimits_{z\in \pi_q} e^{f(z)}}
\end{equation}

The above two strategies both apply multi-layer bidirectional Transformer encoder blocks \cite{vaswani2017attention} concatenated with a deep neural network in the last several layers as the ranking model.


\section{Experiments}
We first describe the settings and hyper-parameter values of the experiment and then summarize the representative results in this section.
\subsection{Experimental Settings}

The value of $k_1$,$k_2$, and $b$ of BM25 is set as 1.2, 200, and 0.75. We fix the value of $\lambda$, $\mu$, and $\sigma$ of LMJM, LMDIR, and LMABS as 0.1, 2000, and 0.7 and use $\delta= 2 $  and $\tau= 0.1$ 
in label adjustment. The value of $\alpha$ and $\beta$ of the frequency-based augmentation is set as 1.7 and 0.75 respectively. We tried training with 10 and 20 negative documents. The method is implemented in PaddlePaddle and we use the official model provided by the organizers for initialization.
Finally, we ensemble the models trained above.
Using different training techniques and strategies, we trained some distinctive debiasing models, and the settings are listed in Section 4.1. After simple linear integration, the final online submission result can be obtained. The performance of each model can be seen in section 4.2.

\subsection{Experimental Results}
We split the validation set into two parts by the query-id, i.e. one set contains approximately 80\% of the whole dataset and the other contains 20\%. Firstly, we select some competitive models based on the DCG@10 on the 80\% validation set, and we present the DCG@10 score of each model in Table \ref{tab:indres}. The first column of the table denotes the name. We trained Transformer with 3 and 12 layers, combined with 10 and 20 negative documents, used different values of batch size (we set the value as 11 and 5 due to the limitation of GPU memory) and tried the training strategy of DLA and fixed propensity, respectively. After linear combination, we got the final score of the above ranking models. The result for the final submission is shown in Table \ref{tab:result}. With our solution, our team $Cannot\ Retrieve$ won the 3rd place with the score of DCG@10 at 9.80. 
\begin{table}
  \caption{Individual scores of the ranking model on the validation set.}
  \label{tab:indres}
  \begin{tabular}{lll}
    \toprule
    Model  & DCG@10 \\
   \cline{2-3}
     & 80\% valid & 20\% valid \\
  
  \midrule 
    12L\_DLA\_neg20\_bs11 &  9.77 & 10.38\\
    12L\_DLA\_neg10\_bs11\_freq &9.75 & 10.34\\ 12L\_DLA\_neg10\_bs11 &  9.88 & 10.56 \\
   3L\_DLA\_neg10\_bs5 &  9.82  & 10.59  \\
   3L\_fixed\_neg10\_bs11\_834proj & 9.81 & 10.65\\
   3L\_fixed\_neg10\_bs11 & 9.82 & 10.66 \\
    \bottomrule
\end{tabular}

\begin{tablenotes} 
\footnotesize
\item[1]
 The ``freq'' indicates that the experiment used the frequency-based augmentation 
 \item[2]  The heuristic-based dense features to hidden vectors and ``834proj'' means we transform the feature into 834 dimensions. If not mentioned specially, the default setting is 869. 
\item[3] The number behind "bs" represents the batch size used in the model.
\end{tablenotes} 
\end{table}
  

\begin{table}
  \caption{Top 5 results for the competition of unbiased learning for web search.}
  \label{tab:result}
  \begin{tabular}{ccl}
    \toprule
    Rank & Team Name & DCG@10\\
    \midrule
    1 & Tencent Search & $10.14328 $ \\
    2 & THUIR & $9.91182$ \\
    \textbf{3} & \textbf{Cannot Retrieve} & \textbf{9.8041 } \\
    4 & Accepted & $7.82751$ \\
    5 & Team of unused99 & $7.06397$ \\
  \bottomrule
\end{tabular}
\end{table}

\section{Conclusion}
In this paper, we present our solution for ``Unbiased learning to Rank'' task of WSDM Cup 2023. We design a feature-enhanced network combined with hybrid debiasing strategies. In the future, we can try to add other feedback signals such as dwelling time to formulate a multi-task learning objective and train on a larger dataset to further enhance the performance. 
\begin{acks}
This work was supported by the Lenovo-CAS Joint Lab Youth Scientist Project. Any opinions, findings, conclusions, and recommendations expressed in this material are those of the authors and do not necessarily reflect those of the sponsors.
\end{acks}

\bibliographystyle{ACM-Reference-Format}
\bibliography{sample-base}


\end{document}